# THE GALAXY GROUP/COSMOLOGY CONNECTIONS

GARY A. MAMON, *IAP, Paris, & DAEC, Obs. de Paris, Meudon, FRANCE*

**Abstract**
Groups of galaxies are highly linked to cosmology: 1) groups are tidally destroyed by the tidal field of the cluster they fall into; 2) spherical infall leads to the young cosmo-dynamical state of loose groups, a fundamental surface for groups, $\Omega_0 \simeq 0.3$, and the mixed nature of compact groups (virialized groups, groups at full collapse, and chance alignments within collapsing loose groups, for decreasing compact group velocity dispersion) ; 3) X-ray analyses lead to $\Omega_0 \lesssim 0.5$.

## 1. Introduction

Whereas clusters of galaxies have often been used to provide cosmological constraints on the Universe, such as the density parameter, $\Omega_0$, and the primordial density fluctuation spectrum, little similar effort has been applied to small groups of say 4 to 30 galaxies, as these suffer from small number statistics in a severe way: their properties (membership, virial $M/L$, dynamical state) are function of the algorithm used to define the groups, and vary tremendously from group to group.

Roughly half of all galaxies lie in groups that are probably bound (*e.g.*, ref. 1]), in contrast with the $\simeq 5\%$ that lie within rich clusters. Much interest has been provoked by the observation of groups that appear very compact in projection on the sky. A well-defined sample of 100 *compact groups* has been generated[2] from visual inspection of POSS plates. These groups appear denser than the cores of rich clusters, and seem now to be the best observed sample of galaxy systems.

Relative to field galaxies, the galaxies in compact groups show a higher level of dynamical interactions[3],[4], ongoing merging[5], and star formation[6]. Also, the morphologies of compact group galaxies are more correlated with group velocity dispersion than with any other group parameter[7],[8],[9],[10], which cannot be explained in simple models where galaxy merging generates elliptical morphologies[11].

Groups are subject to numerous myths, listed below, as should become clear to the reader by the end of this contribution.

*Myth 1:* Groups are virialized.

*Myth 2:* Group dynamics imply $\Omega_0 < 0.1$.

*Myth 3:* X-rays in groups imply $\Omega_0 \simeq 1$.

*Myth 4:* Groups are the preferential site for galaxy evolution.

*Myth 5:* Compact groups are nearly all as dense in 3D as they appear in projection.

## 2. Groups within clusters

The high level of small-scale substructure observed in clusters[12] is usually thought to be caused by groups falling into clusters. The frequency of substructure in clusters is a test on $\Omega_0^{[13],[14],[15]}$, since in a low density universe, structure should freeze out early, and then reach internal equilibrium and become smooth, but one needs to know the dynamical survival time of the substructure, which can be destroyed by the tidal field of the cluster, or merge into other ones. This has been analyzed for X-ray isophotes using dynamical simulations with gas[16]. Simple calculations[17] show that the cluster tidal field is strong enough to destroy infalling groups (except for groups as dense as compact groups appear to be), as is confirmed by more detailed dynamical simulations (Capelato & Mazure 1994 in these proceedings).

The statistics of the primordial density field tell us that small dense systems (high peaks) will form preferentially near large dense ones[18]. Therefore, one expects dense groups to form near rich clusters. Dynamical friction will force these groups to fall to the center of the cluster. The galaxies in the group will merge into ellipticals, either before dynamical friction is completed or afterwards. Such dense groups may thus be the progenitors of a substantial fraction of the elliptical galaxies lying in the cores of clusters[19].

## 3. Spherical infall applied to groups

Groups, as everything else, partake initially in the general Hubble expansion. Because they are selected to be overdense objects, they go through a range of *cosmo-dynamical states*, reaching a maximum expansion (*turnaround*), then collapsing, and finally possibly virializing and/or coalescing into a single galaxy.

## 3.1 How small and how hot must a virialized system be?

From spherical infall, the mean density of a system at turnaround must be[20] $\bar{\rho}_{\text{ta}} = (9\pi^2/16)/(6\pi G t_0^2)$, and that of a system that has virialized must be $\bar{\rho}_{\text{vir}} \geq 8\rho_{\text{ta}}(t_0/\eta) = \Delta_{\text{vir}}/(6\pi G t_0^2)$, where for $\eta = 3$, $\Delta_{\text{vir}} = 81\pi^2/2 \simeq 400$ ($\forall \Omega_0$). Combining this result with the virial theorem, $\sigma_v^2 = \gamma GM/R$, yields a system size $R \leq \sigma_v t_0/(\pi\eta\gamma^{1/2}) = 100\,(\sigma_v/100\,\text{km s}^{-1})h^{-1}$ kpc, for $\eta = 3$, a singular isothermal ($\gamma = 1/2$) and $\Omega_0 = 1$ ($H_0 t_0 = 2/3$). So loose groups (which have $R \geq 1\,h^{-1}$ Mpc) are not virialized. Similarly, one obtains $\sigma_v \geq \gamma^{1/2}(\pi\eta GM/t_0)^{1/3} = 278\,(h\,M_{13})^{1/3}$ km s$^{-1}$ with the same parameters as above. For the typical compact group, $M = 3.8\,h^{-1}10^{12}M_\odot$[21], yielding $\sigma_v \geq 201$ km s$^{-1}$, which happens to be the median compact group velocity dispersion[21], and over twice the median loose group velocity dispersion[1,22].

## 3.2 The fundamental surface

If a group has not yet virialized, one will incorrectly estimate its mass and crossing time. In spherical infall cosmology (and in other cosmologies too) the *biases* in these estimates are well-defined functions of the cosmo-dynamical state. Figure 1a shows *fundamental tracks* for the mass bias versus the crossing time (see ref. [23] for more details on the fundamental surface).

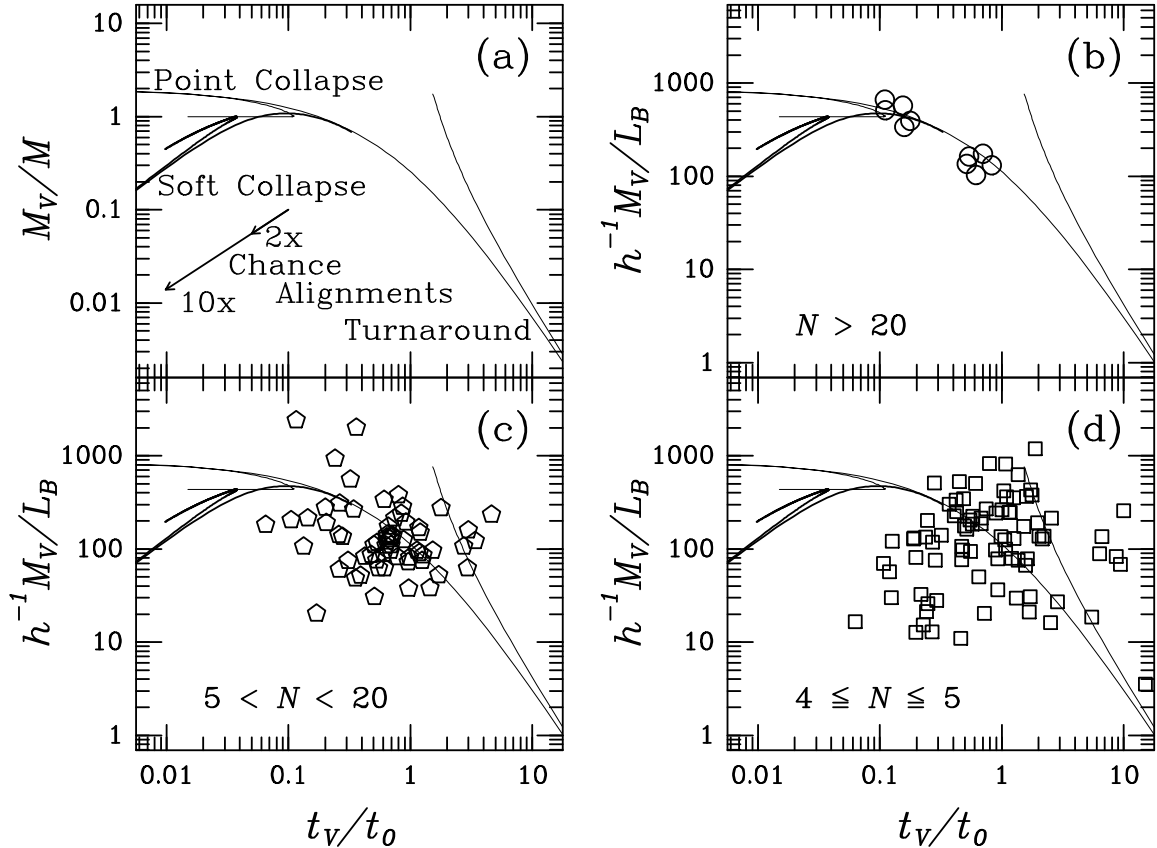

**Figure 1.** The fundamental surface for loose groups, where $t_V$ and $M_V$ are the measured crossing time and virial mass, while $M$ is the true mass.

To compare with observed groups[22], the true $M/L$ of groups is assumed to be a uni-versal constant, hence $M_{\rm vir}/L$ is proportional to the mass bias. Using the observed high multiplicity groups to scale the $y$-axis, one obtains (Fig. 1b) $M/L = 440\,h$, which extrapolated to large scales would yield $\Omega_0 = 0.3$. Figures 1c and 1d show the groups of lower multiplicity, for which the statistical noise on the mass and crossing time estimators becomes increasingly important. The tracks in Figures 1b,c,d are cuts through a non-planar *fundamental surface*, where the third dimension is the scale of the system (*e.g.*, luminosity).

For low $N$, there is an excess of groups below the fundamental track, which is interpreted as favorable projections of elongated prolate groups. Alternatively, this can be caused by enhanced star formation in very small groups (*i.e.*, $M/L \sim N^\alpha$, $\alpha \approx 1/2$, but then many low $N$ groups would still be expanding [upper right track]), or else the spherical infall scenario is far off. Note that only a minority of groups are off below the fundamental track (a negligible number are off above the track, and those are probably unbound groups). And one expects that favorable projections of elongated groups will be more frequent at low multiplicity.

In the standard picture, the low median $M_{\rm vir}/L$ ($100\,h$) is caused by a combination of *near-turnaround bias* and favorable projections. The collapse time of each group can be determined, and from their distribution, one can infer $\Omega_0$ and the primordial density fluctuation spectrum. No loose groups have yet completed their collapse, and the large majority have negligible interpenetration of their galaxy halos. Hence, one does not expect strong environmental effects in loose groups, and if any are present, they may be tracers of effects at galaxy formation.

*3.3 The nature of compact groups*

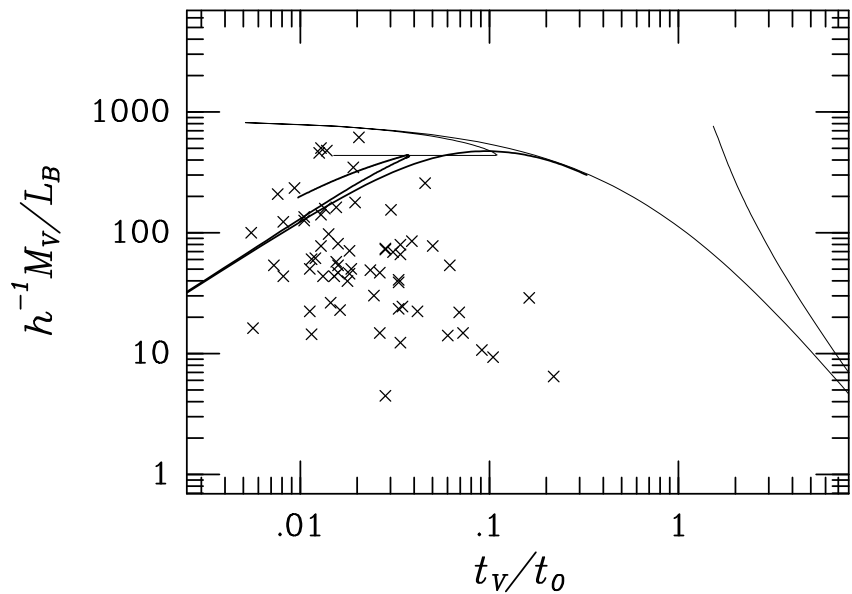

**Figure 2.** The fundamental surface for compact groups

Figure 2 shows that compact groups[21] occupy the lower left part of the fundamental track diagram, because they are selected to be compact, hence small. With this selection criterion, the high velocity dispersion compact groups (upper left) are the ones that are virialized and/or coalescing (see §3.1), next are the groups that are near full collapse, and the low velocity dispersion compact groups (lower right) are caused by chance alignments within *collapsing* loose groups. Alternatively, the low velocity dispersion compact groups could be caused by favorable projections lowering the line-of-sight velocity dispersion, *i.e.*, decreasing $\beta_{\rm spec} = \mu m_p \sigma_v^2/(kT)$, well below the typical values of 0.4. However, first indications[24] yield only few groups with very low $\beta_{\rm spec}$, despite the fact that the *ROSAT* data may select against the very cool groups. Moreover, there is little intergalactic hot gas in low $\sigma_v$ compact groups, but this may simply be an extension of a general $L_X^{\rm IGM}$ vs. $\sigma_v$ relation seen for groups and clusters (Ponman, in these proceedings).

The dynamically cold compact groups account for roughly half of the sample, consistent with the idea that compact groups are mainly caused by chance alignments within larger loose groups[8],[25], in which the high expected frequency of binaries[26] can explain the high level of interactions and star formation (§1). The morphology-velocity dispersion relation in compact groups (§1) merely reflects the fact that the high velocity dispersion compact groups are the only groups that have had enough *time* to reach virialization and see their galaxies merge within them into ellipticals.

## 4. The baryonic fraction in groups of galaxies

The relatively high baryonic fraction (25–30%) in clusters has been used as an argument for a low $\Omega_0$[27]. Similarly, the first observation of intergalactic hot gas in a group (NGC 2300) led to a very low baryonic fraction (4%)[28], consistent with $\Omega_0 = 1$, if extrapolated to large scales. Subsequent studies[29],[30] have yielded baryonic fractions of 10–15% in groups. A reanalysis of the NGC 2300 group[31] points to baryonic fractions of 20 to 30% using the same X-ray surface brightness profile (the discrepancy is caused by the too large background used by the first study), implying $\Omega_0 \lesssim 0.3 \, h_{50}^{-0.92}$ if extrapolated to large scales. Moreover, it has been pointed out[31] that the baryonic fraction increases with radius in the best fitting models, so that deeper X-ray observations may yield even larger numbers. However, the group has been reobserved more deeply, and the baryonic fraction turns out to be of order $\simeq 12\%$ (Mushotzky, private communication), and the discrepancy with the second study would be caused by imperfect galaxy subtraction in the first study! At face value, this last baryonic fraction would extrapolate to $\Omega_0 \simeq 0.5 \, h_{50}^{-0.9}$ on large scales.

*Acknowledgements:* I am indebted to Trevor Ponman for useful discussions, and providing me with a look at his and Harald Ebeling's data, in advance of publication.